# Large Transport Gap Modulation in Graphene via Electric Field Controlled Reversible Hydrogenation

Shaorui Li[1,†], Jiaheng Li[1†], Yongchao Wang[2,†], Chenglin Yu[1], Yaoxin Li[1], Wenhui Duan[1,3,4], Yayu Wang[1,4*], Jinsong Zhang[1,4*]

[1]*State Key Laboratory of Low Dimensional Quantum Physics, Department of Physics, Tsinghua University, Beijing 100084, China*

[2]*Beijing Innovation Center for Future Chips, Tsinghua University, Beijing 100084, China*

[3]*Institute for Advanced Study, Tsinghua University, Beijing 100084, China*

[4]*Frontier Science Center for Quantum Information, Beijing, China*

[†] These authors contributed equally to this work.

* Emails: jinsongzhang@tsinghua.edu.cn; yayuwang@tsinghua.edu.cn

**Abstract**

Graphene is of interest in the development of next-generation electronics due to its high electron mobility, flexibility and stability. However, graphene transistors have poor on/off current ratios because of the absence of a bandgap. One approach to introduce an energy gap is to use hydrogenation reaction, which changes graphene into insulating graphane with $sp^3$ bonding. Here we show that an electric field can be used to control conductor-to-insulator transitions in microscale graphene via a reversible electrochemical hydrogenation in an organic liquid electrolyte containing dissociative hydrogen ions. The fully hydrogenated graphene exhibits a lower limit sheet resistance of 200 G$\Omega\cdot$sq$^{-1}$, resulting in graphene field-effect transistors with on/off current ratios of $10^8$ at room temperature. The devices also exhibit high endurance, with up to one million switching cycles. Similar insulating behaviours are also observed in bilayer graphene, while trilayer graphene remains highly conductive after the hydrogenation. Changes in the graphene lattice, and the transformation from $sp^2$ to $sp^3$ hybridization, is confirmed by in-situ Raman spectroscopy, supported by first-principles calculations.

**Main Text**

Graphene offers high electron mobility, mechanical strength, flexibility and transparency[1]. However, large-area graphene is a semimetal with zero bandgap and the on/off ratio of traditional graphene field-effect transistors (FETs) is low (< 30) at room temperature[2], which is insufficient for logic applications. If the width of the graphene channel is reduced to make a graphene nanoribbon, the quantum confinement effect can create a gap at the Dirac point[3]. However, roughness, geometry, and doping can alter the bandgap of nanoribbon devices[4]. Therefore, very narrow nanoribbons with well-defined edges are required in order to create a useful bandgap for graphene FETs, which creates serious challenges in terms of semiconductor processing[4].

The hydrogenation of graphene can lead to the formation of a stable alkane compound known as graphane (fully hydrogenated graphene). The addition of hydrogen atoms changes the hybridization of the carbon atoms from $sp^2$ to $sp^3$, resulting in a drastically different electronic structure with a large energy gap[5,6]. Controlling the conductor–insulator transition in graphene using a reversible and in situ hydrogenation method is thus a potential alternative method to produce graphene-based logic transistors with sufficient on/off current ratios. However, because of the chemical inertness and weak hydrogen affinity of graphene[7], hydrogenation requires powerful reagents to initiate the chemical reactions, and activation of either hydrogen or the graphene lattice is required to make the reaction feasible[8].

Since its initial synthesis[9], a number of different routes to hydrogenated graphene have been explored. These can be classified into three main approaches: atomic/plasma hydrogen reactions[9–15], chemical reduction[16–18], and electrochemical hydrogenation[19–21]. All these methods produce hydrogenated graphene with insulating behaviour and gap-opening characteristics, but additional processing (such as thermal annealing[9–20], electron-beam irradiation[22] or laser-pulse irradiation[21]) is generally required to restore the pristine graphene, which makes the materials difficult to integrate into FETs. Electric-field-controlled electrochemical gating is a versatile approach since the ion implantation and extraction can be precisely controlled by the gate voltage ($V_G$) [23–29]. Previous attempts with electrochemical graphene FETs used hydrogen ions ($H^+$) from the hydrolysis of residual water in air moisture[23] or organic dielectrics[24]. However, the low concentration of $H^+$ ions limited the switching time, and the high applied gate voltage led to degradation of the graphene devices after many switching cycles.

In this Article, we show that an organic electrolyte can be used in the electric-field-induced reversible hydrogenation of graphene, creating monolayer graphene FETs with on/off ratios of $10^8$ at room temperature. Our hydrogen-ion electrolyte (HIE) has a high concentration of dissociative $H^+$ ions that readily react with graphene, which leads to fast switching speeds and low $V_G$ (between -1 V and 2.6 V) operation within the electrochemical window of our electrolyte. The low $V_G$ avoids irreversible reactions between the anions in the electrolyte and graphene, allowing an endurance of over $10^6$ cycles. We also test the efficacy of our hydrogenation/dehydrogenation process on bilayer and trilayer graphene.

**Electric-field control of hydrogenation in monolayer graphene**

In our experimental setup (Fig. 1a), the graphene devices are immersed in a HIE composed of liquid organic solvent with dissolved $H^+$ ions (the recipe of the HIE is described further in Methods). A thin platinum foil is used as the gate electrode (Supplementary Figure 1). When a positive $V_G$ is applied between the gate and source electrode of graphene, the electric field drives $H^+$ ions to accumulate on top of the graphene surface. Thus, the resistance of graphene is strongly modulated due to the formation of an electric double layer (EDL) at the interface between graphene and HIE. With increasing $V_G$ to above the hydrogenation potential, graphene lattice becomes highly activated and the chemical bonding between $H^+$ ions and C atoms starts to form, as illustrated in Fig. 1b.

Figure 1c displays the drain-source current ($I_{DS}$) of a monolayer graphene (MLG), which reveals a characteristic V-shaped profile[30–32] for -1 V < $V_G$ < 1.5 V arising from the linear dispersion of relativistic Dirac fermions intrinsic to graphene[1,7]. Surprisingly, when $V_G$ is increased to 1.7 V, $I_{DS}$ drops dramatically to nearly zero and remains pinch-off up to $V_G$ = 4 V (Supplementary Figure 2). More transfer curves measured at different temperatures are shown in Supplementary Figure 3. We attribute such a conductor–insulator transition to the hydrogenation of MLG, which opens a wide bandgap analogous to graphane[5,6]. When $V_G$ is swept back from 2.2 V, $I_{DS}$ stays in the "off" state till 0.2 V and then starts to increase gradually. It nearly recovers the original "on" state at -1 V, where the MLG is dehydrogenated. The subsequent two cycles of gate sweeping between -1 V to 2.2 V generate nearly identical $I_{DS}$ traces. Such unprecedented consistency further demonstrates the highly reversible hydrogenation and dehydrogenation processes without the degradation of graphene lattice.

To accurately determine the high resistance value of hydrogenated MLG, we carefully restrain the leakage current ($I_G$) between the gate and graphene electrodes. By linearly fitting the $I_{DS} + I_G$ as a function of $V_{DS}$ (Supplementary Figures 4 and 5), we can derive that the sheet resistance of hydrogenated MLG has a lower limit of 200 $G\Omega \cdot sq^{-1}$ at room temperature (see more discussion in the Supplementary Information).

When the cell is cooled down to 0 °C to further suppress $I_G$, the sheet resistance can reach up to 300 GΩ·sq$^{-1}$ (Supplementary Figure 6), approaching the measurement limit of our instrument. Such high resistance offers an ideal "off" state for graphene-based logic devices. In the "on" state at 1.5 V and -1 V, the MLG is highly conductive with sheet resistance merely ~ 410 Ω·sq$^{-1}$. Therefore, the on/off ratio can reach a record high value of $4.9×10^8$ at room temperature, which is seven orders of magnitude higher than that ever reported on microscale MLG FETs (< 30)[2,4].

The changes of graphene lattice induced by hydrogenation are corroborated by Raman spectroscopy, as shown in Fig. 1d. The main features in the Raman spectra of MLG are the G and 2D peaks at around 1580 and 2670 cm$^{-1}$, resulting from the in-plane vibrational ($E_{2g}$) mode and two-phonon inter-valley triple resonance scattering, respectively[33,34]. The single and sharp 2D peak at charge neutral point ($V_G$ ~ 0.3 V) are unambiguous fingerprints of MLG[33,34]. As more electrons are injected at $V_G$ ~ 1.5 V, the G peak becomes sharper and shifts to 1590 cm$^{-1}$, consistent with previous reports[30,35]. With the further increase of $V_G$ to 2.0 V, the hydrogenation reaction results in three distinctive features in Raman spectra: (1) the appearance of sharp D and D′ peaks at 1340 and 1600 cm$^{-1}$, (2) the significant broadening of the 2D peak, and (3) the emergence of combined (D + D′) and overtone (2D′) modes around 2930 and 3190 cm$^{-1}$. The D peak is undetectable in disorder-free MLG because its activation requires a defect via an intervalley double-resonance Raman process[33,34]. Therefore, the intensity of D peak relative to G peak can serve as a measure for the defect density in graphene. Here, the sharp D peak in hydrogenated graphene reflects the breaking of translational symmetry of C-C sp$^2$ bonds and the formation of massive C-H sp$^3$ bonds[9], behaving like a large amount of defects in graphene lattice. Besides, the observation of D′, D + D′, and 2D′ peaks further proves the presence of defect-assisted activation during the resonance Raman processes. It is worth noticing that the intensity of all Raman peaks are greatly increased compared to that without hydrogenation, which is probably caused by the surface-enhanced Raman scattering[36] induced by the enhanced electromagnetic field between neighbouring isolated and randomly-sized nanoscale pristine graphene domains embedded in the hydrogenated area[18], as further supported by the broad fluorescence background in the Raman spectra (Fig. 1d). The restoration of graphene lattice during the dehydrogenation reaction for $V_G$ from 2 V to -1 V is also confirmed by the Raman spectroscopy (Fig. 1d). When $V_G$ is lowered from 2 V, all the hydrogenation-induced Raman peaks (D, D′, D+D′, and 2D′) gradually decrease in intensity and completely disappear at around -0.5 V, leaving only the G and 2D peaks characteristic of graphene. These observations clearly demonstrate that the hydrogenated graphene can be fully recovered to its original lattice by electric-field control, which is further proved by the highly reversible and reproducible Raman spectra with multicycle sweeps (Supplementary Figure 7). Moreover, the agreement between Raman spectra and transport measurements indicates that the insulating graphene is indeed caused by the hydrogenation-induced transformation from $sp^2$ to $sp^3$ hybridization.

To confirm that the current pinch-off is due to hydrogenation rather than bad connection between drain/source electrodes and graphene flake, we performed the control experiments in a three-segment device on a single graphene ribbon with shared drain/source electrodes as shown in Fig. 2a-c. Only the middle segment is exposed to HIE, while the left and right ones are covered by a layer of 300-nm poly(methyl methacrylate) (PMMA) to obstruct the hydrogenation reaction. We first ground the shared electrode 2 and record the $I_{DS}$ of both the left and middle segments simultaneously. The middle segment reproduces the pinch-off state at $V_G \sim 2$ V (Fig. 2b), while the $I_{DS}$ of the left segment remains "on" at the same $V_G$ (Fig. 2a). This proves that the shared electrode 2 has a good electrical connection with the graphene ribbon even when the middle segment is fully hydrogenated. We then performed similar measurements in the middle and right segments. Likewise, the results in Fig. 2b and 2c completely rule out the possibility that electrode 3 is detached from the graphene ribbon during hydrogenation. It is worth mentioning that $I_{DS}$ of the left and right segments still presents V-shaped behaviour (Fig. 2a and c) despite the PMMA cover layer. This is probably caused by the diffusion of $H^+$ ions from the middle segment via the interface between graphene and PMMA or SiO$_2$ substrate, other than through the PMMA layer (Supplementary Figure 8). The obstruction of hydrogenation in the left and right segments at $V_G \sim 2$ V is attributable to the blockage of the electric field by the PMMA layer, further demonstrating the decisive role of the electric field in controlling the graphene hydrogenation.

The gate-controlled switching capability is another key factor for the application of graphene FETs. As shown in Fig. 2d, when $V_G$ is quickly switched between -0.5 V and 2.4 V, the $I_{DS}$ should exhibit sharp transitions between on and off states, which is explicitly demonstrated by the alternating $V_{DS}$ across graphene channel in the voltage divider circuit (Supplementary Figure 9). After one million switching cycles (Fig. 2e), the graphene FET remains readily functional without any decay of the divided voltage, indicating the high reversibility and stability of the hydrogenation and dehydrogenation in MLG. Moreover, the mobility of graphene is nearly unaffected up to 100k cycles, though decreases gradually to about 2000 cm$^2$V$^{-1}$s$^{-1}$ after 1M cycles as shown in Supplementary Figure 10. All these findings provide a long-sought solution to obtain a giant on/off current ratio and reliable switching capability in graphene FETs, which would significantly advance the applications of graphene logic devices.

Figure 2f and g present the switching speeds of the MLG device under the control of switching $V_G$ between -1 V and 2.6 V. We find that the rise time (10 % − 90 %) of $V_{DS}$ is 0.4 ms and the turn-off delay time is around 2.2 ms, which strongly depends on the turn-off gate voltage and the distance between gate electrode and graphene channel. Thus, this delay time can be attributed to the migration time of H$^+$ ions onto graphene surface driven by the electric field. In Fig. 2g, the fall time to the "on" state is hard to evaluated due to the presence of undesired signal when switching $V_G$ from 2.6 V to -1 V, which probably arises from the large parasitic capacitance in our electrochemical FET. Here we hypothesize the trace of switch curve (blue dashed line) by following the tendency of $V_{DS}$. Thus, the fall time can be estimated to be around 0.13 ms, much lower than the rise time (0.4 ms). And the delay time is around 1.1 ms. This can be interpreted that the strip off of H atoms from graphene should be faster than the migration and accumulation to the graphene surface. Compared with the previous studies on electrochemical graphene FETs[23,24], the performance of our MLG device is much more superior in terms of degradation, response time, and on/off ratio. The key improvement in our devices is utilizing the HIE with dissociative $H^+$ ions other than that from the hydrolysis of residual water in the moister air[23] or organic dielectrics[24]. Moreover, we only applied low gate voltages (between -1 and 2.6 V) to avoid the electrolysis of electrolyte and the irreversible reaction with anions (e.g. $TFSI^-$ or $OH^-$), which may induce strong physical stress and break the C-C bonds in the graphene channel (see more discussions in the Supporting Information).

**Reversible hydrogenation in multilayer graphene**

Besides monolayer graphene, the electric-field-driven reversible hydrogenation in multilayer graphene has also been studied. As shown in Fig. 3a, the $I_{DS}$ of bilayer graphene (BLG) presents a characteristic V-shaped feature[32] similar to MLG before the occurrence of hydrogenation ($V_G$ < 1.5 V). A new feature here is that BLG presents a layer-by-layer hydrogenation behaviour, as reflected by the two-step falloff of $I_{DS}$ for $V_G$ > 1.6 V. Fully hydrogenated BLG is also highly insulating with the sheet resistance larger than 50 GΩ·sq$^{-1}$ (Supplementary Figure 11). As $V_G$ is swept back to -1 V, $I_{DS}$ of hydrogenated BLG progressively recovers its original conducting state. Two subsequent cycles of reversible hydrogenation display reproducible $I_{DS}$ traces (Fig. 3a), with a slight difference in the onset voltages of the second jump at $V_G$ > 1.7 V, which presumably corresponds to the hydrogenation of the bottom layer. The Raman spectra in Fig. 3b further corroborate the electric-field control of reversible hydrogenation in BLG. Near the charge neutral point ($V_G$ ~ 0 V), both the significant broadening and blue-shift of the 2D peak (to 2700 cm$^{-1}$) reveal the fingerprint of BLG. When the top graphene layer is hydrogenated at $V_G$ = 1.7 V, the Raman spectrum exhibits similar changes as that in MLG and remains nearly unchanged during the hydrogenation of bottom layer at $V_G$ = 2.1 V, which indicates consistent hydrogenation mechanism and structure both for the MLG and BLG. As the dehydrogenation is induced by sweeping

$V_G$ to -1 V, the Raman spectrum progressively restores its original characteristics of pristine BLG.

For the trilayer graphene (TLG) with ABA stacking order (Supplementary Figure 12), the $I_{DS}$ traces in Fig. 3c are nearly identical to that of MLG and BLG before the hydrogenation (-1 V < $V_G$ < 1.5 V). However, the hydrogenation of TLG (1.7 – 1.9 V) only decreases the $I_{DS}$ but cannot completely pinch it off. With further increase of $V_G$ to 2.5 V, no more hydrogenation process is detected even up to 3.5 V (Supplementary Figure 13). As $V_G$ is swept back to -1 V, the $I_{DS}$ first decreases monotonically, then fluctuates during the dehydrogenation process and finally merges into the original V-shaped profile. More cyclic electric-field controls of $I_{DS}$ demonstrate the high stability and reversibility of hydrogenation in TLG, as further proved by the representative Raman spectra in Fig. 3d. The consistent Raman results of MLG, BLG, and TLG demonstrate the similar C-H configurations and defect types in the hydrogenated graphene.

**Theoretical explanation of the hydrogenation reaction**

To understand the microscopic mechanism of the electrochemical reaction between hydrogen and graphene, we performed density functional theory (DFT) calculations. Figure 4a displays the potential energy curves for single atomic H on top of the C sites with a fully relaxed MLG lattice as a function of the distance between H and adjacent C atoms. It can be clearly seen that the potential energy curves have a local minimum with the binding energy around 0.8 eV and H-C distance about 1.2 Å. A detailed analysis of the charge density distribution shows that charge transfer occurs between the H atom and graphene lattice, indicating the formation of chemical bond. In this case, the C atom under the H atom moves 0.38 Å upward relative to the graphene plane. Before the formation of H-C bond, there is an energy barrier of ~ 0.2 eV corresponding to the H atom at ~1.8 Å above the C atom, consistent with previous reports[6,37]. When the electric field is applied on top of graphene, a linear background is superposed to the potential energy of the graphene and H atom system. Accordingly, the energy barrier gradually flattens with increasing electric field to ~ 2 V/nm, which is readily achieved in our experiments at $V_G$ = 2 V by the strongly localized electric field of the ~1 nm EDL on graphene surface[31,32]. Moreover, as illustrated in Fig. 4b, the energy barrier is lowered by electron doping to an experimentally accessible level of ~ $1 \times 10^{14}$ cm$^{-1}$ (Ref. 30,32). This finding indicates that the hydrogenation can be greatly facilitated by the strong electric field and high electron doping, in good agreement with our observation that MLG is highly conductive at $V_G$ ~ 1.5 V, which is right before the reaction. For the dehydrogenation process, it is unexpected that MLG gradually becomes conductive with $V_G$ < 0.2 V because the energy barrier of H desorption is as high as 1.0 eV. We attribute such discrepancy to the metastable hydrogenated lattice susceptible to Eley-Rideal reactions between H$^+$ ions in HIE and chemically bonded H atoms[37,38], or the desorption of recombined H atom pairs on graphene surface[39,40].

Among all the structures of fully hydrogenated graphene, the most stable one is supposed to have the chair conformation with the hydrogen atoms alternating on both sides of the plane [5]. Although only the top surface of MLG is exposed to HIE in our devices, we believe that hydrogen atoms are chemically bonded to carbon atoms on both sides based on the following reasons. Firstly, the most stable composition for single-side hydrogenation is $C_4H$ as demonstrated both experimentally and theoretically[11,41–43]. However, the binding energy of $C_4H$ (1.98 eV/H atom)[41] is much smaller than that of double-side hydrogenated chair conformer (6.56 eV/H atom)[5]. Additionally, in the general case with two H atoms bonded to the nearest C atoms, H positions on opposite sides of the graphene sheet are much more favourable than those on the same side[6]. Secondly, the single-side hydrogenated graphene can only be obtained when the hydrogen is completely restricted to one side[41]. Given the high diffusivity of H$^+$ ions under electric field[25–27], they can easily penetrate through the top surface[44,45] or intercalate into the interface between graphene and $SiO_2$ substrate through the exposed edges[46]. Therefore, H atoms prefer to bond both sides of graphene, which is further facilitated by the progressive reduction of formation energies with the increase of H concentrations[6,47]. Thirdly, due to the frustrated hydrogenation domains arising from the random bonding of H atoms to A or B sites of graphene lattice[48], the single-side hydrogenated $C_4H$ with strong disorders is still conductive[11] with sheet resistance much lower than our results. Finally, compared with the plasma-induced single-side hydrogenation[9–11], our Raman spectra are more analogous to that from the chemical reduction method, which has been used to produce high H coverage over 50% (Ref. 16,17).

Based on the above analyses, we employ the chair conformer as the structure of fully hydrogenated MLG and calculate the density of states (DOS). As shown in Fig. 4c, the hydrogenation process converts gapless MLG into an insulator with bandgap up to 3.5 eV. BLG with only the top layer hydrogenated is equivalent to a graphane/graphene (B-A) heterostructure, which is still conductive. Upon complete hydrogenation, the highly insulating BLG possesses a large bandgap corresponding to bilayer graphane (Fig. 4d). For the case of TLG, the applied electric field is strongly localized on the top one/two layers, thus at least one bottom graphene layer cannot be fully hydrogenated and serves as a gapless conducting channel as displayed in Fig. 4e with the graphane/BLG (B-A-A) and bilayer-graphane/MLG (B-B-A) configurations. The electronic structure calculations thus are highly consistent with the transport behaviour of all the devices with different numbers of graphene layers.

**Conclusions**

We have shown that electric-field-induced reversible hydrogenation using a hydrogen ion electrolyte can be used to create a large transport gap in microscale monolayer graphene. The approach can provide graphene FETs with a room-temperature on/off current ratio of $10^8$, high mobility ($> 6,000$ $cm^2V^{-1}s^{-1}$) and switching behaviour that is stable for millions of cycles. This conductor–insulator transition also occurs in bilayer graphene, while trilayer graphene remains conductive. Our technique could potentially be useful in the development of practical graphene-based electronics, such as all-graphene integrated circuits in which monolayer graphene is used to create FETs and trilayer graphene to create interconnection lines[49].

The reversible adsorption of H atoms onto atomically-thin graphene membranes by the electric-field controlled hydrogenation process also represents a feasible approach for hydrogen storage in hydrogenated graphene, which could offer advantages over alternative chemical methods[6,50]. Lastly, and in light of previous studies of $H^+$ ion control in metal oxides and hydrides[25–29], the electric-field-induced hydrogenation could potentially be generalized to other 2D materials. By using it to reversibly tune the atomic bonding and band structures in other low dimensional materials, the in-situ hydrogenation/dehydrogenation method could be used to study their electrical transport and search for new materials.

**Methods**

**Sample preparation**

Graphene crystals were mechanically exfoliated on silicon wafers with 285 nm $SiO_2$. The thickness was identified by the atomic force microscope and Raman spectroscopy. After that, standard electron-beam lithography and thermal evaporation were used to deposit Cr/Au (3/50 nm) on the ends of the graphene samples to serve as the drain and source electrodes. A long strip of Pt foil (5 mm × 2 mm × 25.4 μm) was used as the gate electrode. Then a layer of sealant and cover glass were employed to seal all these electrodes in liquid electrolyte cell as detailed in Ref. 51. The injection of hydrogen ion electrolyte was carried out in argon-filled glovebox and the filling openings were fully sealed by the epoxy before taking it out of the glovebox for the electric transport measurements. The hydrogen ion electrolyte is composed of 0.18 mol/L Bis(trifluoromethane)sulfonamide (HTFSI) in liquid Poly(ethylene glycol) (PEG) with average $M_n$ 600.

**Electric transport measurements**

A direct voltage $V_{DS}$ = 5 mV was applied between the drain and source electrodes by Keithley 2400 SourceMeter, and $I_{DS}$ was recorded simultaneously. Another 2400

SourceMeter was used to apply $V_G$ between the Pt electrode and graphene source electrode and measure $I_G$ in the meanwhile. The change of $V_G$ was performed only at room temperature (~ 25 °C) with the changing rate at 2 mV/s. When measuring the resistance of hydrogenated graphene, $V_G$ was kept unchanged while $V_{DS}$ was swept back and forth with the rate at 5 mV/s. All the sheet resistance values were obtained from the two-terminal resistance by considering the size effect.

**Raman spectroscopy measurements**

Raman spectroscopy was measured by HORIBA LabRAM HR Evolution Spectrometer with 532 nm laser excitation and 1800 l/mm grating. All the Raman spectra have subtracted the background signals contributed by HIE. During Raman measurements, two Keithley 2400 SourceMeters worked together to control $V_G$ and monitor $I_{DS}$ respectively. When $V_G$ reached the target voltage, the Raman signals were acquired until $I_{DS}$ became nearly stable.

**Theoretical calculations**

First-principles calculations were performed in the framework of density functional theory using the Vienna ab initio Simulation Package[52]. The energy cutoff of the plane-wave basis was fixed at 500eV. The projector augmented wave (PAW) method and the Perdew-Burke-Ernzerhof type generalized gradient approximation (GGA)[53] were adopted to model the electron-ion interactions and the exchange correlations between electrons, respectively. The crystal structure relaxations were performed with the force criteria of 0.01 eV/Å. The van der Waals interaction between neighbouring layers was considered by the DFT-D3 methods[54]. A vacuum layer of 20 Å was added in the direction perpendicular to 2D planes to avoid interactions between adjacent periodic images. The Monkhorst-Pack k-point meshes of 40×40×1, 9×9×1 were adopted for pristine lattices and supercells of 3×3, respectively.

**Data availability:**

The data that support the findings of this study are available from the corresponding author upon reasonable request.

## References

1. Castro Neto, A. H., Guinea, F., Peres, N. M. R., Novoselov, K. S. & Geim, A. K. The electronic properties of graphene. *Rev. Mod. Phys.* **81**, 109–162 (2009).

2. Novoselov, K. S. *et al.* Electric Field Effect in Atomically Thin Carbon Films. *Science* **306**, 666–669 (2004).

3. Li, X., Wang, X., Zhang, L., Lee, S. & Dai, H. Chemically Derived, Ultrasmooth Graphene Nanoribbon Semiconductors. *Science* **319**, 1229–1232 (2008).

4. Schwierz, F. Graphene transistors. *Nat. Nanotechnol.* **5**, 487–496 (2010).

5. Sofo, J. O., Chaudhari, A. S. & Barber, G. D. Graphane: A two-dimensional hydrocarbon. *Phys. Rev. B* **75**, 153401 (2007).

6. Boukhvalov, D. W., Katsnelson, M. I. & Lichtenstein, A. I. Hydrogen on graphene: Electronic structure, total energy, structural distortions and magnetism from first-principles calculations. *Phys. Rev. B* **77**, 035427 (2008).

7. Novoselov, K. S. *et al.* A roadmap for graphene. *Nature* **490**, 192–200 (2012).

8. Whitener, K. E. Review Article: Hydrogenated graphene: A user's guide. *J. Vac. Sci. Technol. A* **36**, 05G401 (2018).

9. Elias, D. C. *et al.* Control of Graphene's Properties by Reversible Hydrogenation: Evidence for Graphane. *Science* **323**, 610–613 (2009).

10. Luo, Z. *et al.* Thickness-Dependent Reversible Hydrogenation of Graphene Layers. *ACS Nano* **3**, 1781–1788 (2009).

11. Son, J. *et al.* Hydrogenated monolayer graphene with reversible and tunable wide band gap and its field-effect transistor. *Nat. Commun.* **7**, 1–7 (2016).

12. Chen, H. *et al.* Fabrication of Millimeter-Scale, Single-Crystal One-Third-Hydrogenated Graphene with Anisotropic Electronic Properties. *Adv. Mater.* **30**, 1801838 (2018).

13. Ryu, S. *et al.* Reversible Basal Plane Hydrogenation of Graphene. *Nano Lett.* **8**, 4597–4602 (2008).

14. Bostwick, A. *et al.* Quasiparticle Transformation during a Metal-Insulator Transition in

Graphene. *Phys. Rev. Lett.* **103**, 056404 (2009).

15. Balog, R. *et al.* Bandgap opening in graphene induced by patterned hydrogen adsorption. *Nat. Mater.* **9**, 315–319 (2010).

16. Yang, Z., Sun, Y., Alemany, L. B., Narayanan, T. N. & Billups, W. E. Birch Reduction of Graphite. Edge and Interior Functionalization by Hydrogen. *J. Am. Chem. Soc.* **134**, 18689–18694 (2012).

17. Yang, Y., Li, Y., Huang, Z. & Huang, X. (C1.04H)n: A nearly perfect pure graphane. *Carbon* **107**, 154–161 (2016).

18. Schäfer, R. A. *et al.* On the Way to Graphane—Pronounced Fluorescence of Polyhydrogenated Graphene. *Angew. Chem. Int. Ed.* **52**, 754–757 (2013).

19. Daniels, K. M. *et al.* Evidences of electrochemical graphene functionalization and substrate dependence by Raman and scanning tunneling spectroscopies. *J. Appl. Phys.* **111**, 114306 (2012).

20. Zhao, M., Guo, X.-Y., Ambacher, O., Nebel, C. E. & Hoffmann, R. Electrochemical generation of hydrogenated graphene flakes. *Carbon* **83**, 128–135 (2015).

21. Zhong, Y. L. & Swager, T. M. Enhanced Electrochemical Expansion of Graphite for in Situ Electrochemical Functionalization. *J. Am. Chem. Soc.* **134**, 17896–17899 (2012).

22. Lee, W.-K., Whitener, Jr., Keith E., Robinson, J. T. & Sheehan, P. E. Patterning Magnetic Regions in Hydrogenated Graphene Via E-Beam Irradiation. *Adv. Mater.* **27**, 1774–1778 (2015).

23. Echtermeyer, T. J. *et al.* Nonvolatile Switching in Graphene Field-Effect Devices. *IEEE Electron Device Lett.* **29**, 952–954 (2008).

24. Hayashi, C. K., Garmire, D. G., Yamauchi, T. J., Torres, C. M. & Ordonez, R. C. High On-Off Ratio Graphene Switch via Electrical Double Layer Gating. *IEEE Access* **8**, 92314–92321 (2020).

25. Lu, N. *et al.* Electric-field control of tri-state phase transformation with a selective dual-ion switch. *Nature* **546**, 124–128 (2017).

26. Ji, H., Wei, J. & Natelson, D. Modulation of the Electrical Properties of VO2 Nanobeams Using an Ionic Liquid as a Gating Medium. *Nano Lett.* **12**, 2988–2992 (2012).

27. Tan, A. J. *et al.* Magneto-ionic control of magnetism using a solid-state proton pump. *Nat. Mater.* **18**, 35 (2019).

28. Kremers, M. *et al.* Optical transmission spectroscopy of switchable yttrium hydride films. *Phys. Rev. B* **57**, 4943–4949 (1998).

29. Huiberts, J. N. *et al.* Yttrium and lanthanum hydride films with switchable optical properties. *Nature* **380**, 231–234 (1996).

30. Das, A. *et al.* Monitoring dopants by Raman scattering in an electrochemically top-gated graphene transistor. *Nat. Nanotechnol.* **3**, 210–215 (2008).

31. Chen, F., Qing, Q., Xia, J., Li, J. & Tao, N. Electrochemical Gate-Controlled Charge Transport in Graphene in Ionic Liquid and Aqueous Solution. *J. Am. Chem. Soc.* **131**, 9908–9909 (2009).

32. Ye, J. *et al.* Accessing the transport properties of graphene and its multilayers at high carrier density. *Proc. Natl. Acad. Sci.* **108**, 13002–13006 (2011).

33. Malard, L. M., Pimenta, M. A., Dresselhaus, G. & Dresselhaus, M. S. Raman spectroscopy in graphene. *Phys. Rep.* **473**, 51–87 (2009).

34. Ferrari, A. C. & Basko, D. M. Raman spectroscopy as a versatile tool for studying the properties of graphene. *Nat. Nanotechnol.* **8**, 235–246 (2013).

35. Pisana, S. *et al.* Breakdown of the adiabatic Born–Oppenheimer approximation in graphene. *Nat. Mater.* **6**, 198–201 (2007).

36. Wang, Z., Wu, S., Ciacchi, L. C. & Wei, G. Graphene-based nanoplatforms for surface-enhanced Raman scattering sensing. *Analyst* **143**, 5074–5089 (2018).

37. Sha, X. & Jackson, B. First-principles study of the structural and energetic properties of H atoms on a graphite (0001) surface. *Surf. Sci.* **496**, 318–330 (2002).

38. Sha, X., Jackson, B. & Lemoine, D. Quantum studies of Eley–Rideal reactions between H atoms on a graphite surface. *J. Chem. Phys.* **116**, 7158–7169 (2002).

39. Hornekær, L. *et al.* Metastable Structures and Recombination Pathways for Atomic Hydrogen on the Graphite (0001) Surface. *Phys. Rev. Lett.* **96**, 156104 (2006).

40. Hornekær, L. *et al.* Clustering of Chemisorbed H(D) Atoms on the Graphite (0001) Surface due to Preferential Sticking. *Phys. Rev. Lett.* **97**, 186102 (2006).

41. Li, Y. & Chen, Z. Patterned Partially Hydrogenated Graphene (C4H) and Its One-Dimensional Analogues: A Computational Study. *J. Phys. Chem. C* **116**, 4526–4534 (2012).

42. Haberer, D. *et al.* Evidence for a New Two-Dimensional C4H-Type Polymer Based on Hydrogenated Graphene. *Adv. Mater.* **23**, 4497–4503 (2011).

43. Boukhvalov, D. W. & Katsnelson, M. I. Chemical functionalization of graphene. *J. Phys.: Condens. Matter* **21**, 344205 (2009).

44. Hu, S. *et al.* Proton transport through one-atom-thick crystals. *Nature* **516**, 227–230 (2014).

45. Lozada-Hidalgo, M. *et al.* Sieving hydrogen isotopes through two-dimensional crystals. *Science* **351**, 68–70 (2016).

46. Bediako, D. K. *et al.* Heterointerface effects in the electrointercalation of van der Waals heterostructures. *Nature* **558**, 425 (2018).

47. Stojkovic, D., Zhang, P., Lammert, P. E. & Crespi, V. H. Collective stabilization of hydrogen chemisorption on graphenic surfaces. *Phys. Rev. B* **68**, 195406 (2003).

48. Flores, M. Z. S., Autreto, P. A. S., Legoas, S. B. & Galvao, D. S. Graphene to graphane: a theoretical study. *Nanotechnology* **20**, 465704 (2009).

49. Lin, Y.-M. *et al.* Wafer-Scale Graphene Integrated Circuit. *Science* **332**, 1294–1297 (2011).

50. Subrahmanyam, K. S. *et al.* Chemical storage of hydrogen in few-layer graphene. *Proc. Natl. Acad. Sci.* **108**, 2674–2677 (2011).

51. Zhang, J. *et al.* Reversible and selective ion intercalation through the top surface of few-layer MoS 2. *Nat. Commun.* **9**, 5289 (2018).

52. Kresse, G. & Furthmüller, J. Efficient iterative schemes for ab initio total-energy calculations using a plane-wave basis set. *Phys. Rev. B* **54**, 11169–11186 (1996).

53. Perdew, J. P., Burke, K. & Ernzerhof, M. Generalized Gradient Approximation Made Simple. *Phys. Rev. Lett.* **77**, 3865–3868 (1996).

54. Grimme, S., Antony, J., Ehrlich, S. & Krieg, H. A consistent and accurate ab initio parametrization of density functional dispersion correction (DFT-D) for the 94 elements H-Pu. *J. Chem. Phys.* **132**, 154104 (2010).

**Acknowledgments:**

We thank Yong Xu for helpful discussions and technical supports. This work is supported by the Basic Science Center Project of NSFC (grant No. 51788104) and the National Key R&D Program of China (grants No. 2018YFA0307100 and 2016YFA0301001). This work is supported in part by the Beijing Advanced Innovation Center for Future Chip (ICFC).

**Author contribution statement**

J.S.Z. and Y.Y.W. proposed and supervised the research. J.S.Z designed the device structure and proposed the electrolyte. S.R.L., Y.C.W., C.L.Y., and Y.X.L. fabricated the devices and carried out the electric measurements. S.R.L and Y.C.W. measured the Raman spectra. W.H.D. and J.H.L. performed theoretical calculations. J.S.Z., Y.Y.W., and S.R.L. prepared the manuscript with comments from all authors.

**Competing interests:**

The authors declare no competing interests.

**Figures and captions**

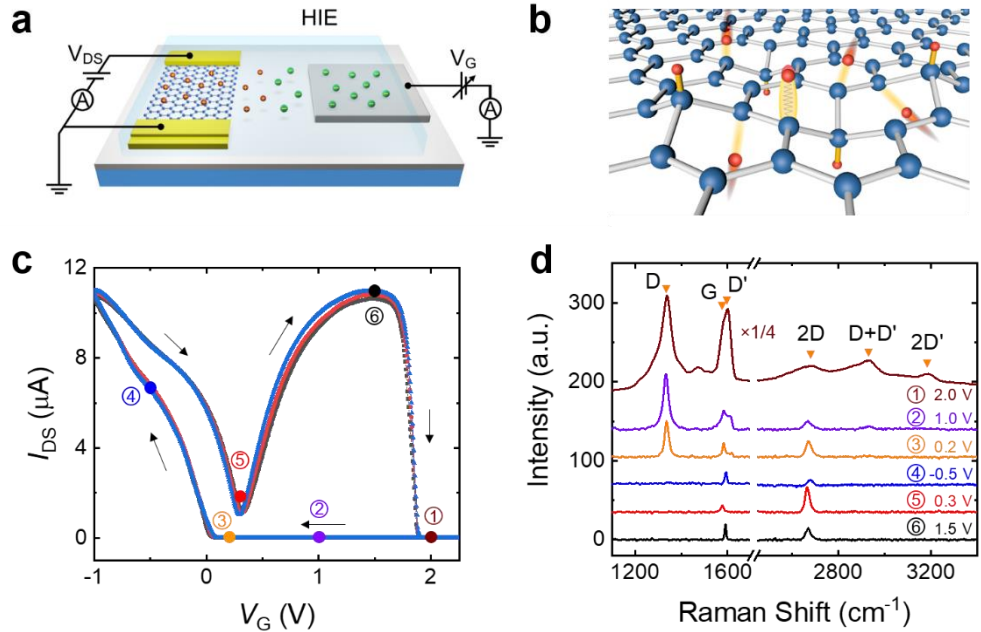

**Fig.1 | Electric-field control of reversible hydrogenation in MLG. a**, Schematic structure of HIE gated graphene device and electric measurement configuration. **b**, Illustration of the hydrogenation process between graphene lattice (blue) and H$^+$ ions (red). **c**, Three consecutive cycles of $I_{DS}$ measurements in device M1 as a function of applied gate voltage $V_G$ at room temperature (25 °C). It represents a highly reproducible conductor-insulator transition in MLG. The black arrows indicate the sweep directions of $V_G$. The applied $V_{DS}$ is fixed at 5 mV. **d**, Raman spectra of MLG at different $V_G$'s in device M2. The corresponding $I_{DS}$ of labelled $V_G$ is represented by the solid dot and circled number with the same colour in **c**. The correspondence of $I_{DS}$ and Raman spectra at different $V_G$'s is highly reproducible in different devices. To clearly present all the Raman spectra, the intensity of $V_G$ = 2.0 V is divided by four.

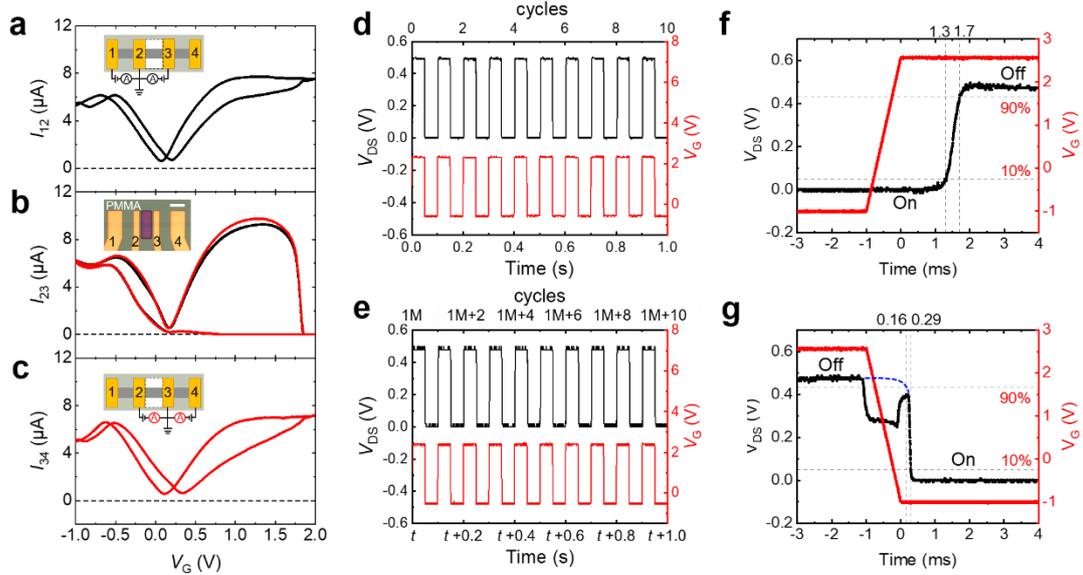

**Fig.2 | Characterization of the insulating and switching behaviours of hydrogenated MLG**. **a**, The current between electrode 1 and 2, $I_{12}$, versus $V_G$ in the three-segment device M3 as shown in the inset. The middle segment is exposed to HIE, whereas the left and right ones are covered by a layer of 300-nm PMMA. **b**, The current between electrodes 2 and 3, $I_{23}$, as a function of $V_G$. The black curve is obtained simultaneously with that in **a**. The inset shows the optical image of our device and the scale bar is 10 μm. **c**, The current between electrodes 3 and 4 measured simultaneously with the red curve in **b**. The measurement configuration is displayed in the inset. **d**, The $V_{DS}$ across the drain and source electrodes of MLG device M4 (black line, left scale) and the square waveform of $V_G$ (red line, right scale) measured by the oscilloscope simultaneously. When $V_G$ is periodically changed between -0.5 V and 2.4 V, the $V_{DS}$ can be quickly switched accordingly in the voltage divider circuit with the input voltage of 0.5 V. The period of $V_G$ is 0.1 s. **e**, Same measurements as **d** after one million switching cycles with $t \sim 28$ hours. The $V_{DS}$ remain unchanged as the beginning, which indicates the high stability of reversible hydrogenation in MLG under the electric-field control. **f-g**, The characterization of respond time in MLG device M4. The rise time of $V_{DS}$ (10 % - 90 %) from on to off state is 0.4 ms and the fall time is estimated around 0.13 ms. The $V_G$ is switched between -1 V and 2.6 V with the changing time of 1 ms. The blue dashed line in **g** is the extrapolated trace of $V_{DS}$ from the tendency.

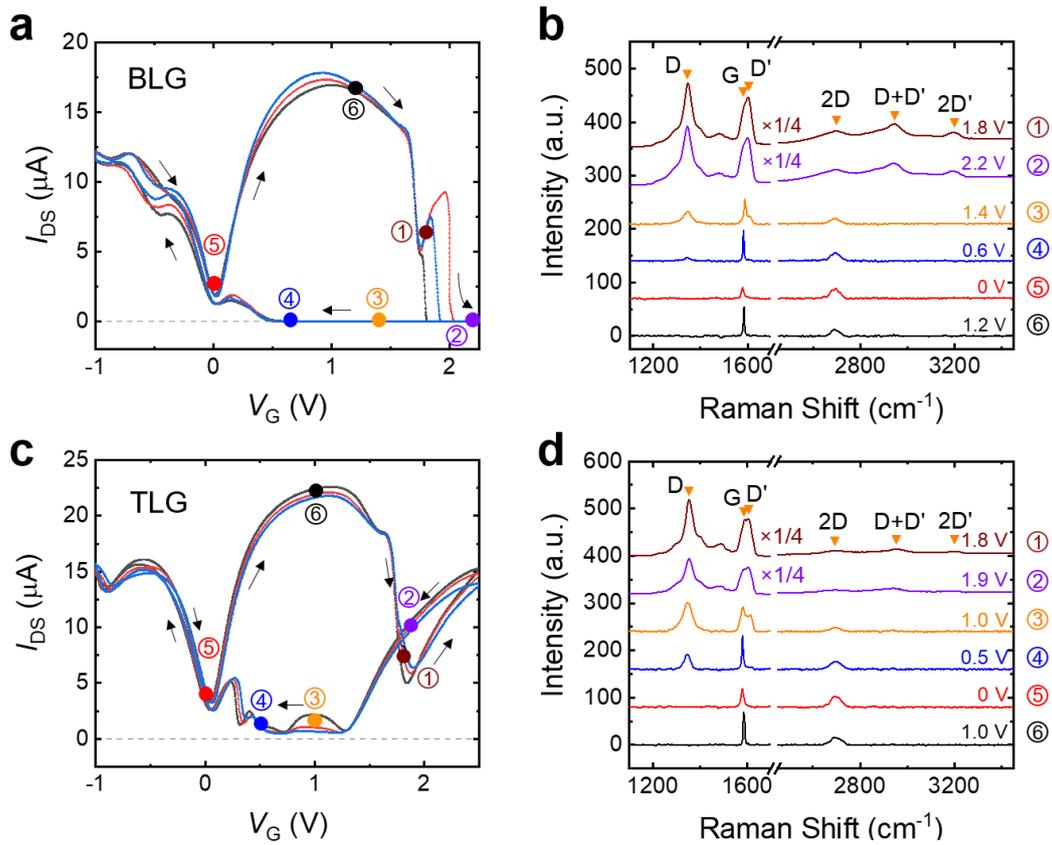

**Fig.3 | Electric-field control of reversible hydrogenation in BLG and TLG. a**, Three consecutive cycles of $I_{DS}$ in BLG as a function of applied gate voltage $V_G$ of device B1. The black arrows indicate the sweep directions of $V_G$. The applied $V_{DS}$ is fixed at 5 mV. **b**, Raman spectra of BLG at different $V_G$'s in device B2. The corresponding $I_{DS}$ of labelled $V_G$ is represented by the solid dot and circled number with the same colour in **a**. The intensity of Raman spectra for $V_G$ = 1.8 and 2.2 V is divided by four. **c-d,** The same measurements of $I_{DS}$ (**c**) and Raman spectra (**d**) in TLG devices (T1 and T2).

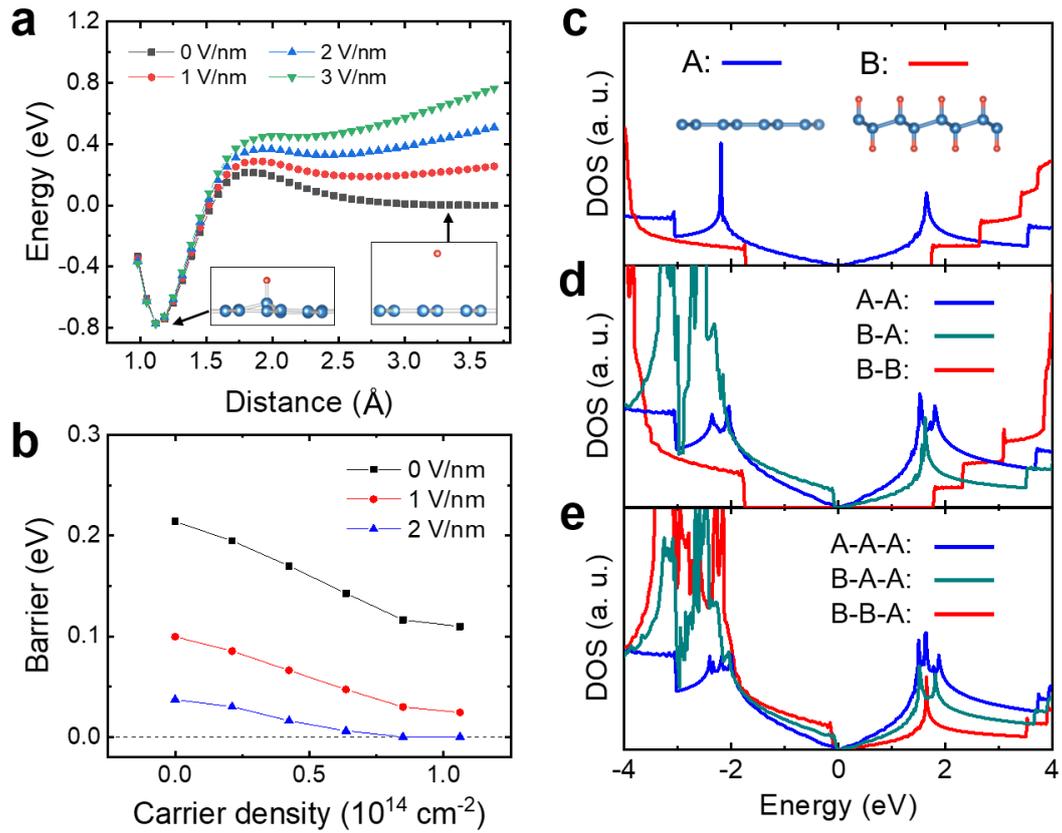

**Fig.4 | DFT calculations of hydrogenated MLG, BLG, and TLG. a,** Potential energy as a function of the distance between H atom and the subjacent C atom under different electric fields. The insets present the configurations of H atom (red) and graphene lattice (blue) with different distances. The effect of electric fields is assumed to add linear backgrounds to the potential energy curves. The barrier becomes nearly flattened with the increase of the electric field up to 2 V/nm. **b,** The energy barrier as a function of doped electron density with different electric fields. **c,** Density of states of MLG (blue) and graphane with chair conformation (red). The crystal structures of single-layer graphene (A) and graphane (B) are presented in the insets. **d-e,** DOS of different stacking configurations in the bilayer (**d**) and trilayer cases (**e**). The stacking sequences are indicated in the top right legends.